\documentclass[a4paper]{jpconf}
\usepackage{graphicx}

\begin{document}
\title{Detecting solar g modes with ASTROD}

\author{ R Burston$^1$, L Gizon$^1$, T Appourchaux$^2$, W-T Ni$^{3,4}$\\
and the ASTROD I ESA cosmic vision 2015-2025 team}

\address{$^1$ Max-Planck-Institut f\"ur Sonnensystemforschung, 37191 
Katlenburg-Lindau, Germany\\}
\address{$^2$ Institut d'Astrophysique Spatiale, Centre universitaire d'Orsay, 91405 Orsay Cedex, France\\}
\address{$^3$ Purple Mountain Observatory, Chinese Academy of Sciences, Nanjing, 210008, China\\}
\address{$^4$ National Astronomical Observatories, Chinese Academy of Sciences, Beijing, 100012, China\\}
\ead{burston@mps.mpg.de}

\begin{abstract}
We present an up-to-date estimate for the prospect of using the {\it Astrodynamical Space Test of Relativity using Optical Devices} (ASTROD) \cite{Ni2006a,Ni2006,Ni2007,Paton2007} for an unambiguous 
detection of solar g modes ($\,f\,<400\,\,\mu$Hz) through their 
gravitational signature. There are currently two major efforts to 
detect low-frequency gravitational effects, ASTROD and the Laser Interferometer 
Space Antenna (LISA) \cite{LISA2000}. Using the most recent g mode surface amplitude estimates, both observational and theoretical, it is unclear whether LISA will be capable of successfully detecting 
these modes. The ASTROD project may be better suited for detection as its sensitivity curve is shifted towards lower frequencies with the best sensitivity occurring in the range $100-300\,\,\mu$Hz.
\end{abstract}

\section{Introduction}

For the past 20 years, traditional helioseismology techniques have provided a wealth of information regarding the internal structure and dynamics of the Sun in both the convection and radiation zones \cite{Gough2006}.  Except for the solar core, probing the Sun's interior has been achieved by direct analysis of the pressure modes detected in Doppler observations of surface radial velocities.  

However, the g modes (or gravity modes), that probe the very core of the Sun where thermonuclear reactions take place, remain elusive. These modes are expected to have very small velocity amplitudes at the solar surface, thereby making them very difficult to detect using conventional Doppler techniques.  For a review of the search of g modes using Doppler velocity data see  \cite{Elsworth2006} and references therein. Furthermore, Garcia et al. \cite{Garcia2007} recently claimed to have detected a collective signature of the dipole g modes. 

In order to make significant progress towards a total understanding of the internal structure and dynamics of the Sun, from the solar surface down through the core to the very center, we require an unambiguous detection of solar g modes.  One of the main goals of the ASTROD project  is to search for the signature of g modes in the gravitational field of the Sun \cite{Ni2006a,Ni2007,Paton2007}.

\section{ASTROD I, II and III}

\begin{figure}[h]
\begin{center}
\includegraphics[width=4.715in]{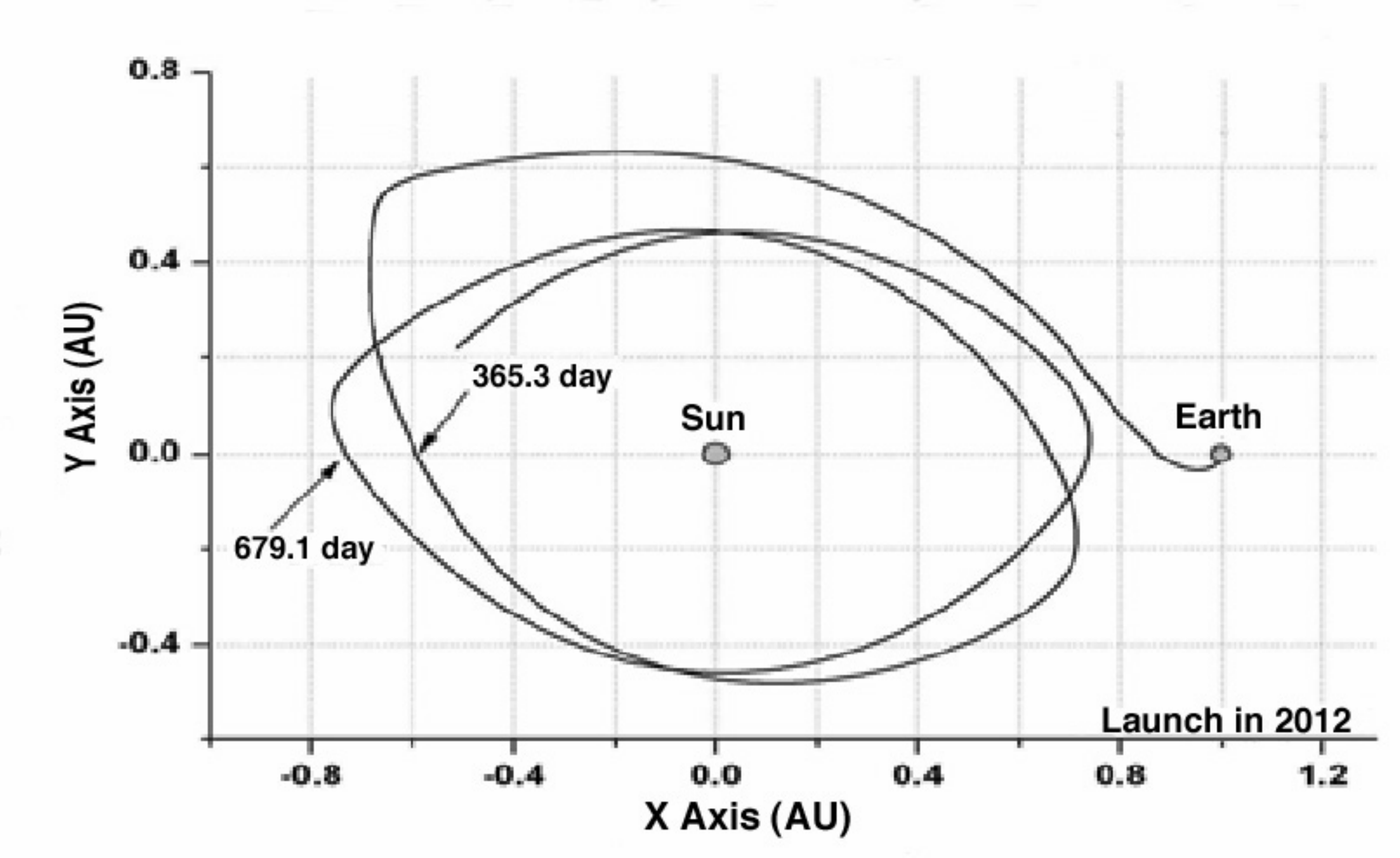}\caption{The 2012 orbit in the Sun-Earth fixed frame.}\label{astrodI}
\end{center}
\end{figure}

The primary objective for the long-term ASTROD concept is to maintain a minimum of 3 spacecraft in orbit within our solar system using laser interferometric ranging to ultimately test relativity. This is separated into 3 distinct stages, each with increasing orders of scientific benefits and engineering milestones. The first stage, ASTROD I, will comprise a single spacecraft in communication with ground stations using laser pulse ranging, orbiting the Sun at an average distance of approximately 0.6 AU. See figure \ref{astrodI} for a schematic of the proposed orbit design.

\begin{figure}[h]
\begin{center}
\includegraphics[width=4.715in]{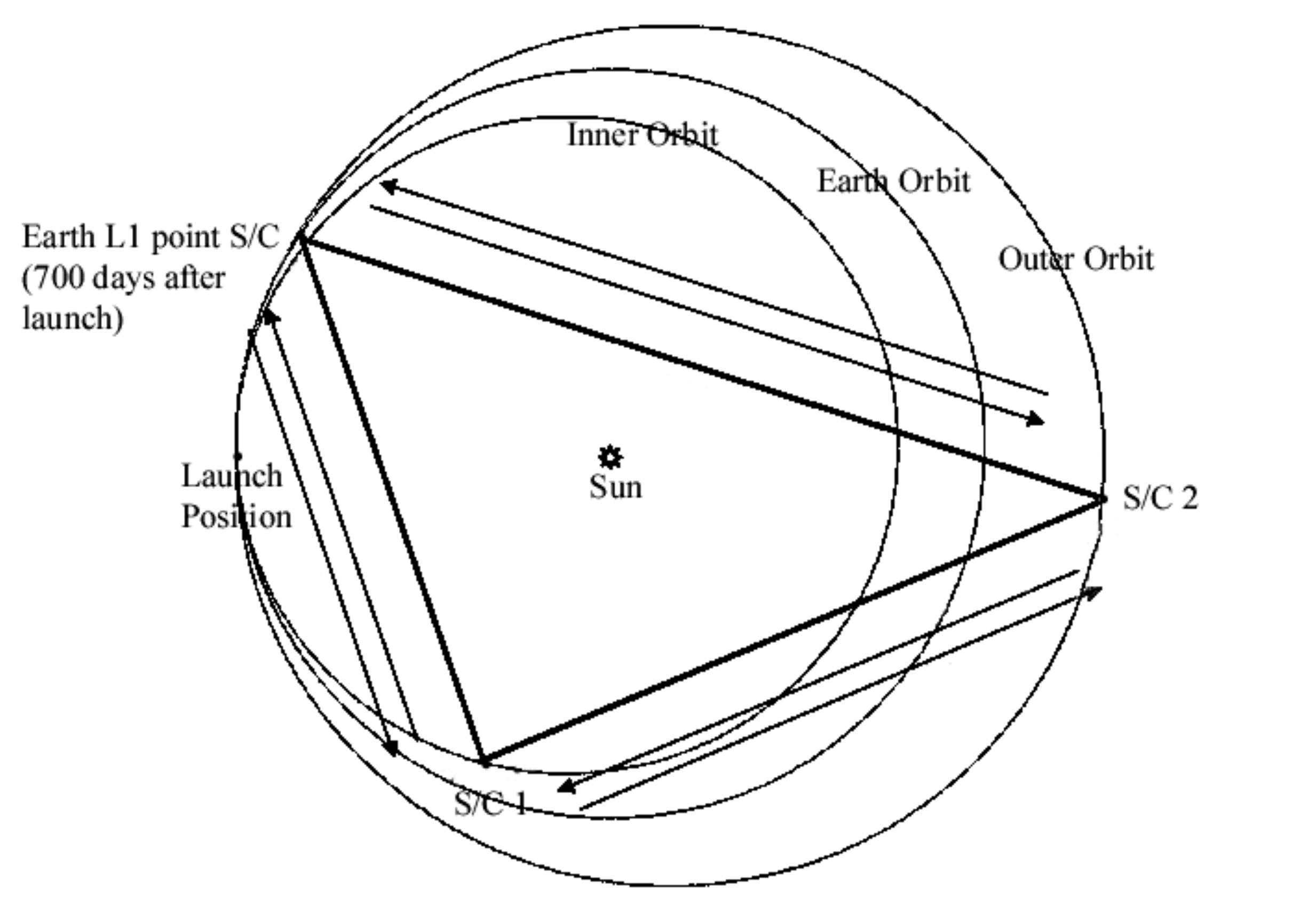}
\caption{A schematic ASTROD II configuration 700 days from launch \cite{Ni2006a}.}\label{astrodII}
\end{center}
\end{figure}

The second phase, ASTROD II, will consist of three spacecraft communicating with 1-2 Watt CW lasers. Two of these (S/C 1 and S/C 2) will be in separate solar orbits and a third (S/C) is to be situated at the Sun-Earth Lagrangian point L1 (or L2). It is this configuration which is capable of detecting gravitational effects induced by solar g modes. See figure \ref{astrodII} for a depiction of the orbit design and configuration of the spacecraft 700 days after launch.

The third and final stage, ASTROD III or Super ASTROD, will then explore the possibility of larger orbits in an effort to detect even lower frequency primordiial waves.

\section{Up-To-Date detectability estimate for g modes}

\begin{figure}[h]
\begin{center}
\includegraphics[width=5in]{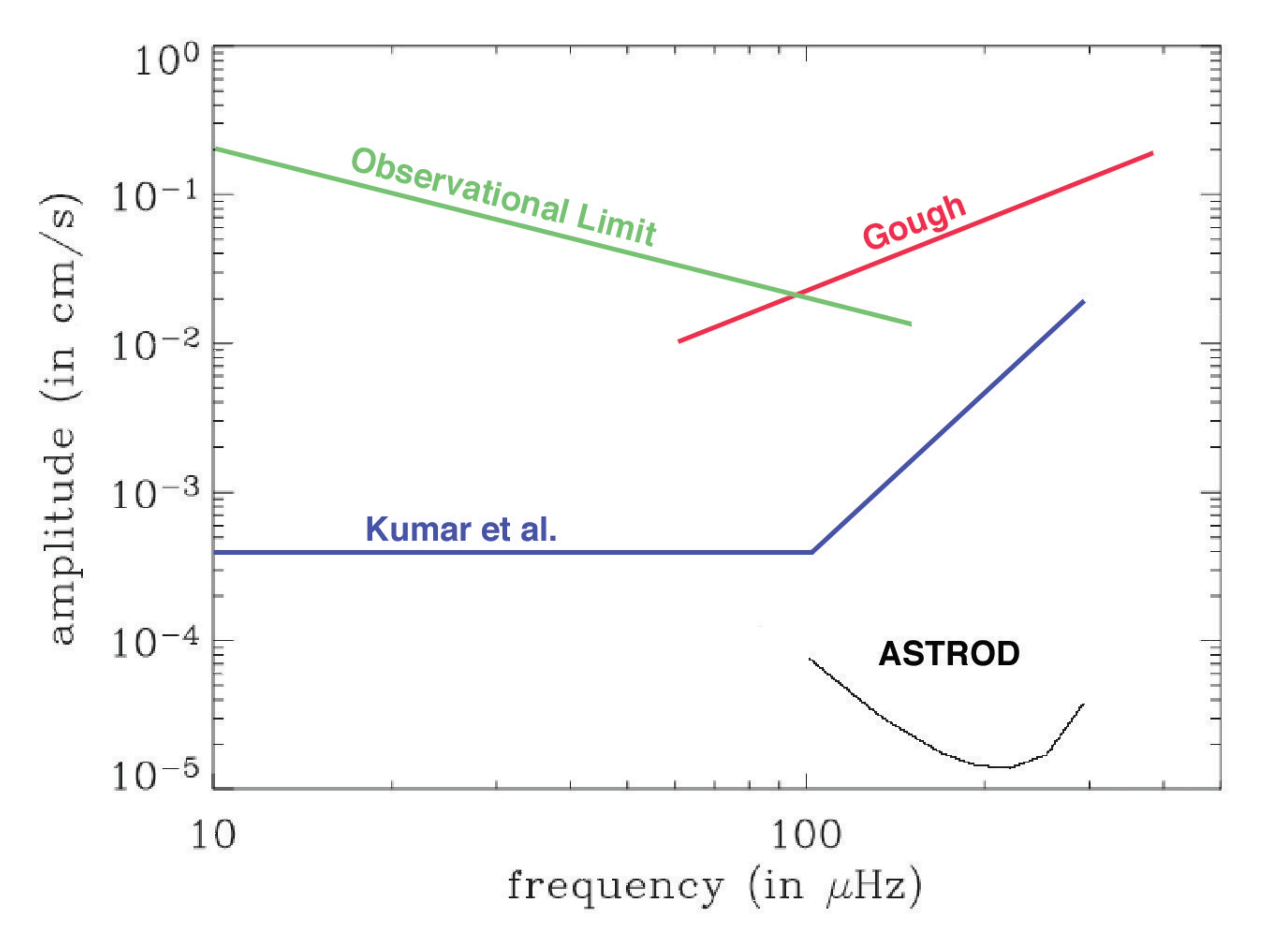}
\caption{Comparison of surface radial velocity amplitudes for $\ell=2$ g modes (quadrupole modes): Theoretical estimates reproduced from Gough \cite{Gough1985} (red curve) and Kumar et al. \cite{Kumar1996}  (blue curve); The green line is an observational limit that was estimated from Garcia et al. \cite{Garcia2007}; The black curve is the ASTROD detection level assuming a one year integration time and a spacecraft orbiting at 0.4 AU.}
\label{astrodplotdsa}
\end{center}
\end{figure}

Figure \ref{astrodplotdsa} displays various, theoretical and observed, limitations for the solar velocity amplitudes of the quadrupole ($\ell=2$) g modes, as well as the anticipated gravity strain sensitivities for ASTROD.

The two theoretical models comprise the most pessimistic estimate from Kumar et al. \cite{Kumar1996} (blue curve) and the most optimistic estimate from Gough \cite{Gough1985} (red curve); see \cite{JCD2002} for a similar discussion. Both Gough and Kumar et al. assume that the g-modes are stochastically excited. The paper by Gough presents an application of the energy equipartition result of Goldreich and Keeley \cite{Goldreich1977a,Goldreich1977b}. Whereas Kumar et al. use the theoretical work of Goldreich et al. \cite{Goldreich1994} and show that the model predicts p mode amplitudes which match observations very well and are therefore confident that any errors in their calculations will be less than one order of magnitude. 

The observational upper-limit is continuously being improved with significant advances in both data analysis methods and data acquisition technology. Reliable estimates were published in 2000 by Appouchaux et al. \cite{Appourchaux2000}, whereby they analysed 784 days of SOHO/MDI data. They found that the $\ell=2$ quadrupole g mode amplitudes are no greater than approximately 9 mm s$^{-1}$ at 200 $\mu$Hz, which decreases to 2.5 mm s$^{-1}$ at 1000 $\mu$Hz; this can be roughly approximated to display a $1/\nu$ dependence as mentioned by \cite{JCD2002}. Subsequently, the Pheobus group \cite{Appourchaux2003} analysed 9 years of BiSON data to reduce the limit to around 3.5 mm $s^{-1}$ at 300 $\mu$Hz which corresponds to about 5 mm s$^{-1}$ at 200 $\mu$Hz by again using the rough $1/\nu$ dependence. Here, the upper-limit is reduced further again to  1 mm $s^{-1}$ at 200 $\mu$Hz. This was estimated from the noise levels in Garcia et al. \cite{Garcia2007} and this corresponds to an average of 50 modes observed by the GOLF instrument with 10 years of data (green curve).

The surface velocity amplitudes for ASTROD were calculated using the most recent gravity strain sensitivities in Paton et al. \cite{Paton2007}, the equations presented in Cutler et al. \cite{Cutler1996} and taking into consideration that we have 2-3 spacecraft rather than a signal interferometer. The gravitational strain falls off as $1/d^4$ where $d$ is the distance to the Sun.  ASTROD's capacity to detect g modes is due to a combination of good strain sensitivity, small distance from the Sun and large distance between spacecraft.

\section{Discussion}

By assuming the best possible sensitivity for ASTROD ($10^{-23}$ at 100 $\mu$Hz) and even the most pessimistic amplitudes of Kumar et al., ASTROD may deliver unambiguous detection for g modes with frequencies between 100 $\mu$Hz and 300 $\mu$Hz.  Moreover, the detection of such ``high frequency'' g modes should provide better diagnostics of the inside of the Sun than the lower frequency g modes.


\section{References}

\begin{thebibliography}{13}


\bibitem{Ni2006a} Ni W-T et al. 2006 {\it J. Phys: Conf. Series} {\bf 32} 154-60

\bibitem{Ni2006} Ni W-T et al. 2006 {\it Acta Astronautica} {\bf 59}  598

\bibitem{Ni2007} Ni W-T 2007 {\it Nuclear Physics B (Proc. Suppl.)} {\bf 166} 153

\bibitem{Paton2007} Pulido-Paton A and Ni W-T 2007 {\it Preprint arXiv:0704.3333v1 [astro-ph]}; General Relativity and Gravitation, in press.

\bibitem{LISA2000} LISA 2000 {\it ESA System and Technology Study Report}, ESA-SCI 11

\bibitem{Gough2006} Gough D O 2006 {\it Proc. of SOHO-17, 10 Years of SOHO and Beyond (Sicily)} (ESA SP-617), Lacoste H and Ouwehand L eds.,   published on CDROM p 1.1

\bibitem{Elsworth2006} Elsworth Y et al. 2006 {\it Proc. of SOHO 18/GONG 2006/HELAS I, Beyond the spherical Sun (Sheffield)} (ESA SP-624), Fletcher K and Thompson M eds, published on CDROM p 22.1

\bibitem{Garcia2007} Garcia R A et al. 2007 {\it May Science express report}

\bibitem{Appourchaux2000} Appourchaux T et al. 2000 {\it ApJ} {\bf 538} 401-14

\bibitem{Appourchaux2003} Appourchaux T 2003 {\it Proc. of SOHO 12/GONG+ 2002} 131-8

\bibitem{Kumar1996} Kumar P et al. 1996 {\it ApJ Letters} {\bf 458} L83

\bibitem{Gough1985} Gough D O 1985 {\it in Future mission in solar, heliospheric and space plasmas physics (ESA SP-235)}, E. Rolfe and B. Battrick eds, 183

\bibitem{JCD2002} Christensen-Dalsgaard J 2002 {\it International Journal of Modern Physics D} {\bf 11} No. 7, 995-1009
\bibitem{Goldreich1977a} Goldreich P and Keeley D 1977 {\it ApJ} {\bf 211} 934-42 

\bibitem{Goldreich1977b} Goldreich P and Keeley D 1977 {\it ApJ} {\bf 212} 243-51 

\bibitem{Goldreich1994} Goldreich P, Murray N and Kumar P {\it ApJ} {\bf 424} 466-79
 
\bibitem{Cutler1996} Culter C and Lindblom L 1996 {\it Phys. Rev. D} {\bf 54} 1287


\end{thebibliography}
\end{document}